\RequirePackage{lineno}
\documentclass[twocolumn,showpacs,superscriptaddress,amsmath,amssymb]{revtex4-1}
\usepackage{graphicx}
\usepackage{epsfig}
\usepackage{dcolumn}
\usepackage{bm}
\usepackage{overpic}
\usepackage{color}
\usepackage{multirow}
\usepackage{subfigure}

\newcommand{\ee}{e^{+}e^{-}~\rightarrow~\Lambda_{c}^{+} \bar{\Lambda}_{c}^{-}}
\newcommand{\Lam}{\Lambda_{c}}
\newcommand{\lam}{\Lambda_{c}^{+}}
\newcommand{\kst}{K_{S}^{0}}
\newcommand{\lambar}{\bar{\Lambda}_{c}^{-}}

\newcommand{\lamdecay}{\Lambda_{c}^{+}\rightarrow pK^{-}\pi^{+}}

\newcommand{\ebm}{E_{\textrm{beam}}}
\newcommand{\mbc}{M_{\textrm{BC}}}
\newcommand{\dele}{\Delta E}

\newcommand{\gev}{\mathrm{GeV}}
\newcommand{\mev}{\mathrm{MeV}}
\newcommand{\mevcc}{\mathrm{MeV}/c^{2}}
\newcommand{\EE}{e^{+}e^{-}}

\newcommand{\modeI}{ pK^{-}\pi^{+}}
\newcommand{\modeII}{ pK_{S}^{0}}
\newcommand{\modeIII}{ \Lambda\pi^{+}}
\newcommand{\modeIV}{ pK^{-}\pi^{+}\pi^{0}}
\newcommand{\modeV}{ pK_{S}^{0}\pi^{0}}
\newcommand{\modeVI}{ \Lambda\pi^{+}\pi^{0}}
\newcommand{\modeVII}{ pK_{S}^{0}\pi^{+}\pi^{-}}
\newcommand{\modeVIII}{ \Lambda\pi^{+}\pi^{+}\pi^{-}}
\newcommand{\modeIX}{ \Sigma^{0}\pi^{+}}
\newcommand{\modeX}{ \Sigma^{+}\pi^{+}\pi^{-}}

\newcommand{\LLB}{\Lambda_{c}^{+}\bar{\Lambda}_{c}^{-}}

\newcommand{\ra}{\rightarrow}
\newcommand{\sqs}{\sqrt{s}}
\newcommand{\ratios}{|G_{E}/G_{M}|}

\newcommand{\alpHaA}{ -0.13\pm0.12\pm0.08 }
\newcommand{\alpHaD}{ -0.20\pm0.04\pm0.02 }
\newcommand{\RatioA}{ 1.14\pm0.14\pm0.07 }
\newcommand{\RatioD}{ 1.23\pm0.05\pm0.03 }

\newcommand{\BornCSA}{ 236\pm11\pm46 }
\newcommand{\BornCSB}{ 207\pm17\pm13 }
\newcommand{\BornCSC}{ 245\pm19\pm16 }
\newcommand{\BornCSD}{ 237\pm^{{\color{white}1}}3^{{\color{white}}}\pm15 }

\newcommand{\ENERGYAT}{ 4574.5 }

\newcommand{\ENERGYBT}{ 4580.0 }

\newcommand{\ENERGYCT}{ 4590.0 }

\newcommand{\ENERGYDT}{ 4599.5 }

\begin{document}
\title{\boldmath Precision measurement of the $\ee$ cross section near threshold}
\author{
\begin{small}
M.~Ablikim$^{1}$, M.~N.~Achasov$^{9,d}$, S. ~Ahmed$^{14}$, M.~Albrecht$^{4}$, M.~Alekseev$^{55A,55C}$, A.~Amoroso$^{55A,55C}$, F.~F.~An$^{1}$, Q.~An$^{52,42}$, J.~Z.~Bai$^{1}$, Y.~Bai$^{41}$
% ...
\\
\vspace{0.2cm}
(BESIII Collaboration)\\
\vspace{0.2cm} {\it
$^{1}$ Institute of High Energy Physics, Beijing 100049, People's Republic of China\\
% ...
$^{4}$ Bochum Ruhr-University, D-44780 Bochum, Germany\\
% ...
$^{9}$ G.I. Budker Institute of Nuclear Physics SB RAS (BINP), Novosibirsk 630090, Russia\\
% ...
$^{14}$ Helmholtz Institute Mainz, Johann-Joachim-Becher-Weg 45, D-55099 Mainz, Germany\\
% ...
$^{41}$ Southeast University, Nanjing 211100, People's Republic of China\\
$^{42}$ State Key Laboratory of Particle Detection and Electronics, Beijing 100049, Hefei 230026, People's Republic of China\\
% ...
$^{52}$ University of Science and Technology of China, Hefei 230026, People's Republic of China\\
% ...
$^{55}$ (A)University of Turin, I-10125, Turin, Italy; (B)University of Eastern Piedmont, I-15121, Alessandria, Italy; (C)INFN, I-10125, Turin, Italy\\
% ...
$^{d}$ Also at the Novosibirsk State University, Novosibirsk, 630090, Russia\\
% ...
}
\vspace{0.4cm}
}
\noaffiliation{}

\date{\today}

\begin{abstract}
The cross section of the $\ee$ process is measured with unprecedented precision using data collected with the BESIII detector at $\sqs=\ENERGYAT$, $\ENERGYBT$, $\ENERGYCT$ and $\ENERGYDT$~$\mev$. The non-zero cross section near the $\LLB$ production threshold is cleared. 
At center-of-mass energies $\sqs=\ENERGYAT$ and $\ENERGYDT$~$\mev$, the higher statistics data enable us to measure the $\Lam$ polar angle distributions. From these, the $\Lam$ electric over magnetic form factor ratios ($\ratios$) are measured for the first time. They are found to be $\RatioA$ and $\RatioD$ respectively, where the first uncertainties are statistical and the second are systematic. 
\end{abstract}

\maketitle

The electromagnetic structure of hadrons, parameterized in terms of electromagnetic form factors (EMFFs), provides a key to understand quantum chromodynamics effects in bound states. The nucleon has been studied rigorously for more than 60 years, but new techniques and the availability of data with larger statistics from modern facilities have given rise to a renewed interest in the field, \textit{e.g.} the proton radius puzzle~\cite{pradius}. Recently, the access to strange and charm hyperon structure by time-like EMFFs provides an additional dimension. Assuming that one-photon exchange dominates the production of spin-1/2 baryons $B$, the cross section of the process $\EE$~$\ra$~$B\bar{B}$ can be parameterized in terms of EMFFs,  \textit{i.e.} $G_{E}$ and $G_{M}$, in the following way~\cite{OPEXModel}:
\begin{equation}
\sigma_{B\bar{B}}(s)=\frac{4\pi\alpha^{2} C \beta}{3s}|G_{M}(s)|^{2}\big[1+\frac{2m_{B}^{2}c^{4}}{s}\big|\frac{G_{E}(s)}{G_{M}(s)}\big|^{2}\big].
\label{OnePhotonPre}
\end{equation}
Here, $\alpha$ is the fine-structure constant, $\beta$=$\sqrt{1-4m_{B}^{2}c^{4}/s}$ the velocity of the baryon, $s$ the square of the center-of-mass (CM) energy, and $m_{B}$ is the mass of the baryon. The Coulomb factor $C$ parameterizes the electromagnetic interaction between the outgoing baryon and antibaryon. 
For neutral baryons, the Coulomb factor is unity, while for point-like charged fermions it reads $C=\varepsilon R$~\cite{C1,C2}, where $\varepsilon = {\pi\alpha}/{\beta}$ is an enhancement factor resulting in a nonzero cross section at threshold and $R={\sqrt{1-\beta^{2}}}/(1-e^{-\pi\alpha{\sqrt{1-\beta^{2}}}/\beta})$ is the Sommerfeld resummation factor~\cite{C1}. The ratio of EMFFs associated with the polar angle distribution of the baryon can also parameterize the differential production cross section of the corresponding baryon~\cite{OPEXModel}.

In the $\EE\ra p\bar{p}$ process, the BaBar collaboration observed a rapid rise of the cross section near threshold, followed by a plateau around 200 MeV above threshold~\cite{babar2}. The BESIII collaboration also observed the cross section enhancement~\cite{ppbarbes}. The non-vanishing cross section near threshold as well as the wide-range plateau have led to various theoretical interpretations, including i) final-state interactions~\cite{TheoFSInter}, ii) bound states  or meson-like resonances~\cite{TheoRes} and iii) an attractive Coulomb interaction~\cite{Resumfct}. Recently, the BESIII collaboration has observed the non-zero cross section near threshold in the process $\EE\ra\Lambda\bar{\Lambda}$~\cite{BAM0139}.  Naturally, it is also interesting to explore the production behavior of $\lam$, the lightest baryon containing the charm quark. Previously, the Belle collaboration measured the cross section of $\ee$ using the initial-state radiation (ISR) technique~\cite{Belle}, but the results suffer from significant uncertainties in CM energy and cross section. Therefore, near $\lam\lambar$ threshold, precise measurements of the production cross section and EMFF ratios are highly desirable.

In this work, the cross section of the reaction $\ee$ is measured at four CM energies: $\sqs=\ENERGYAT$, $\ENERGYBT$, $\ENERGYCT$, and $\ENERGYDT$ $\mev$. At each CM energy, ten Cabibbo-favored hadronic decay modes, $\lam\ra\modeI$, $\modeII$, $\modeIII$, $\modeIV$, $\modeV$, $\modeVI$, $\modeVII$, $\modeVIII$, $\modeIX$, and $\modeX$, as well as the ten corresponding charge-conjugate modes are independently used to reconstruct $\lam$ or $\lambar$. Each mode will produce one measurement of the cross section and the total cross section is obtained from a weighted average over the 20 individual measurements. In addition, the higher statistics data samples at $\sqs=\ENERGYAT$ and $\ENERGYDT$~$\mev$ enable the study of the polar angle distribution of $\Lam$ in the CM system. From these distributions, the ratios between the electric and the magnetic form factors, \textit{i.e.} $\ratios$, are extracted for the first time.

The data samples are collected with the BESIII detector~\cite{BESIII} at BEPCII. The detector has a geometrical acceptance of 93\% of the 4$\pi$ solid angle. It contains a small-celled, helium-based main drift chamber (MDC), a time-of-flight system (TOF) based on plastic scintillators, an electromagnetic calorimeter (EMC) made of CsI(Tl) crystals, a muon system (MUC) made of Resistive Plate Chambers, and a superconducting solenoid magnet. 

Monte Carlo (MC) simulations based on {\sc geant4}~\cite{geant4} are performed to determine detection efficiencies, optimize selection criteria, extract signal shapes and study backgrounds. The $\EE$ collisions are simulated by the {\sc kkmc} generator~\cite{KKMC}, which takes the beam energy spread and the ISR correction into account. The distribution of the $\Lam$ polar angle is considered in the generator by parameterizing it with the function $f(\theta)~\propto~1~+~\alpha_{\Lam}~\cos^2\theta$. After an iterative procedure, the values of $\alpha_{\Lam}$ at $\sqs$~=~$\ENERGYAT$ and $\ENERGYDT$~$\mev$ are obtained from real data (see Table~\ref{AngDisParas}) and at the remaining CM energies by a linear interpolation. 

Using the branching fractions (BR) measured in Ref.~\cite{BAM0162}, all tagged $\Lam$ decays are simulated by weighting phase-space events according to the decay behavior observed in real data. The subsequent decays listed by the particle data group (PDG)~\cite{PDG} are modeled with {\sc evtgen}~\cite{EVTGEN}. The inclusive MC samples include $\lam\lambar$ pair production, $\ell^{+}\ell^{-}$~($\ell=\textit{e,~$\mu$,~$\tau$})$ events, open charm processes~\cite{DDsMC}, ISR-produced low-mass $\psi$ states and the continuum process $\EE \ra q\bar{q}~(q=\textit{u,~d,~s})$.  

Charged tracks as well as the intermediate states $\pi^{0}$, $\kst$, $\Lambda$, $\Sigma^{0}$ and $\Sigma^{+}$ are selected and reconstructed with the same method described in Ref.~\cite{BAM0162}.

In the final states of decay modes $\modeV$ and $\modeVII$, potential background from $\Lambda \rightarrow p\pi^{-}$ is eliminated by rejecting events with $M_{p\pi^{-}}$ lying in the mass window (1100,~1125)~$\mevcc$, where $M_{p\pi^{-}}$ is the invariant mass of $p\pi^{-}$ combinations in the final state. For the decay mode $\modeX$, the corresponding exclusion window is (1110,~1120)~$\mevcc$ due to the smaller observed width of the $M_{p\pi^{-}}$ peak in data. Similarly, background from the intermediate state $\Sigma^{+}$ is removed from the $\modeV$ sample by rejecting events with $M_{p\pi^{0}}$ in the mass window (1170,~1200)~$\mevcc$. In modes $\modeVIII$ and $\modeX$, events with $M_{\pi^{+}\pi^{-}}$ within (490,~510)~$\mevcc$ are rejected to suppress $\kst$ background.

\begin{figure}[!htbp]
\setlength{\abovecaptionskip}{0.2cm}
\setlength{\belowcaptionskip}{-0.5cm}
\begin{center}
\begin{overpic}[width=1.68in,height=1.12in,angle=0]{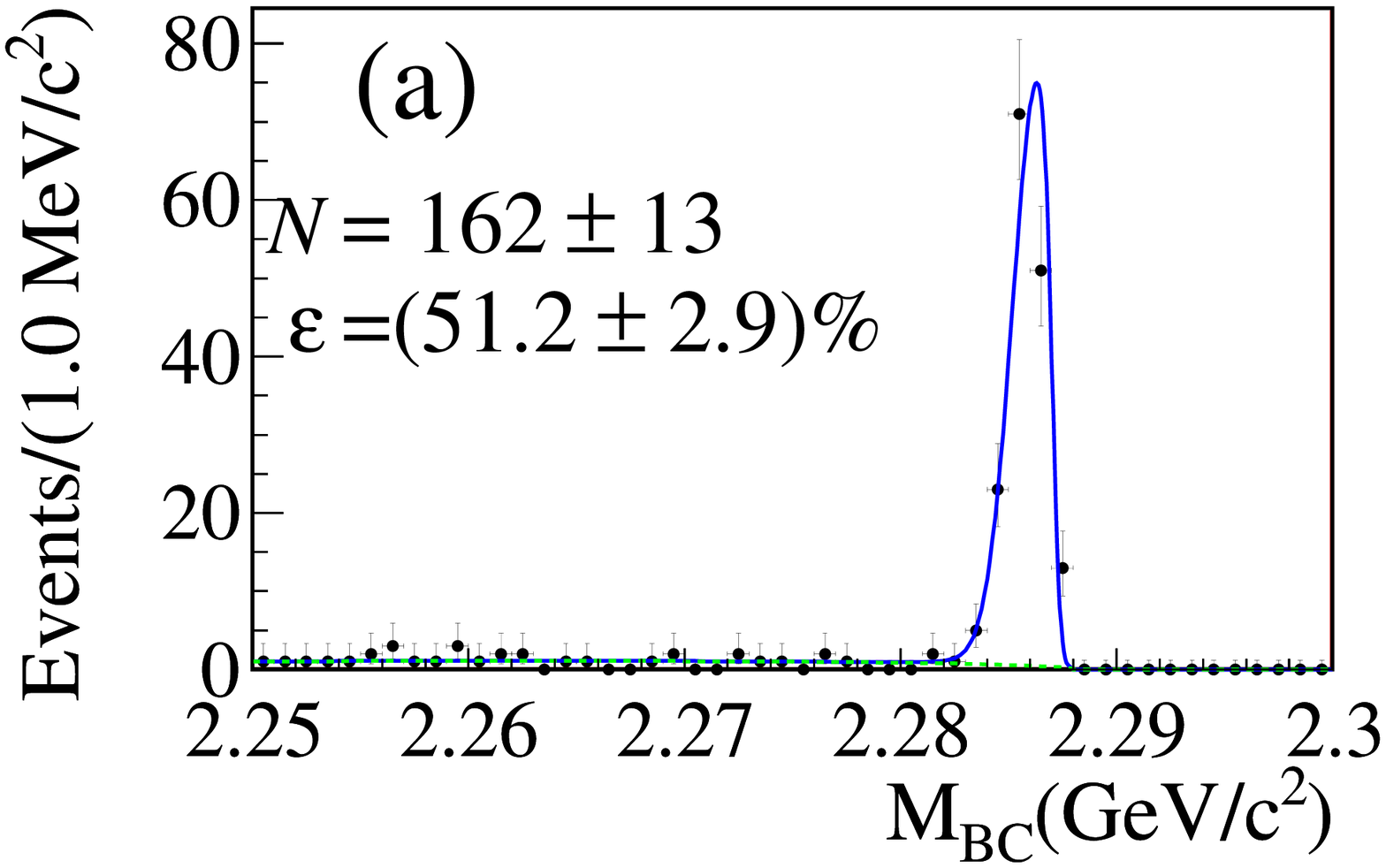}
\end{overpic}
\begin{overpic}[width=1.68in,height=1.12in,angle=0]{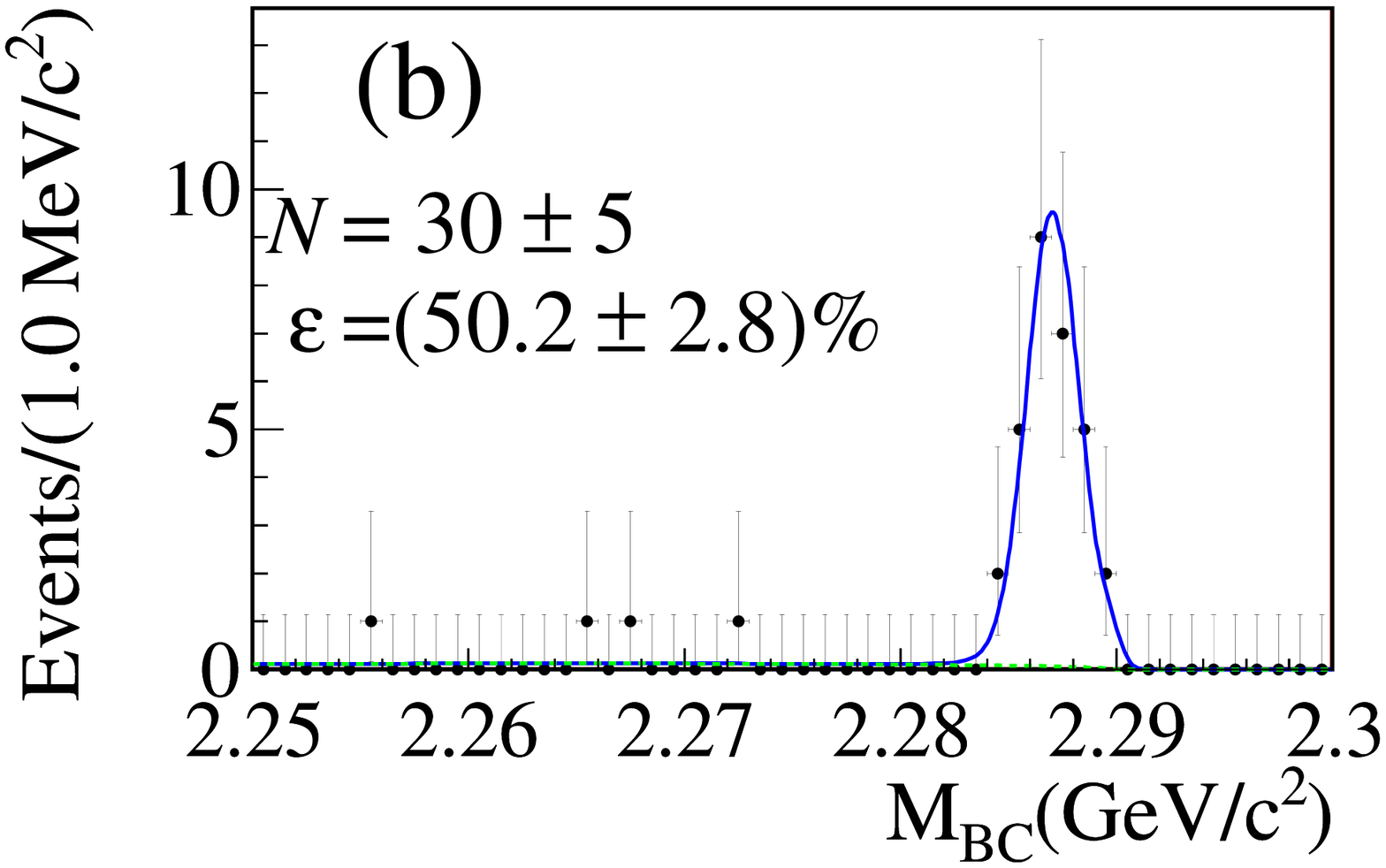}
\end{overpic}
\begin{overpic}[width=1.68in,height=1.12in,angle=0]{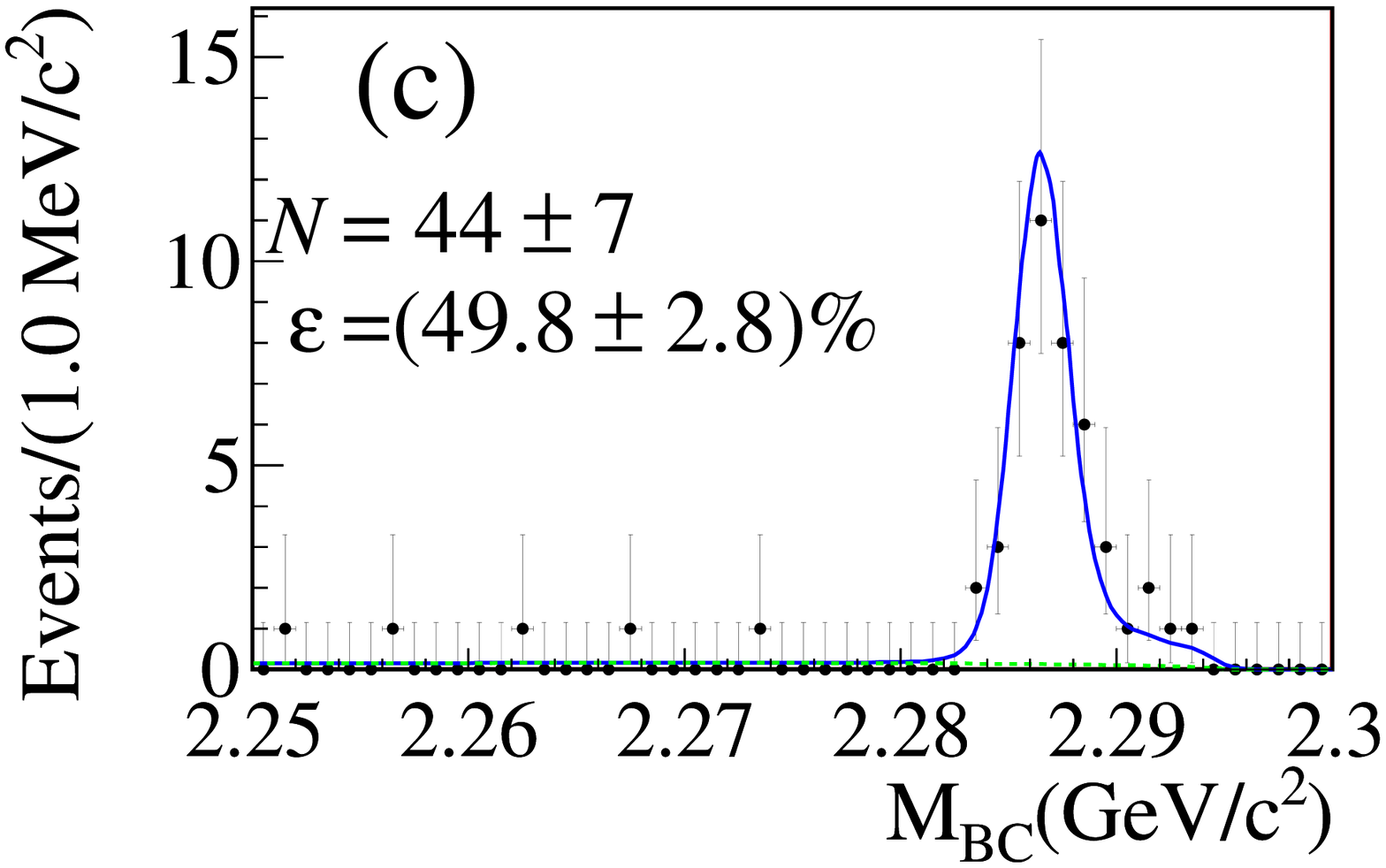}
\end{overpic}
\begin{overpic}[width=1.68in,height=1.12in,angle=0]{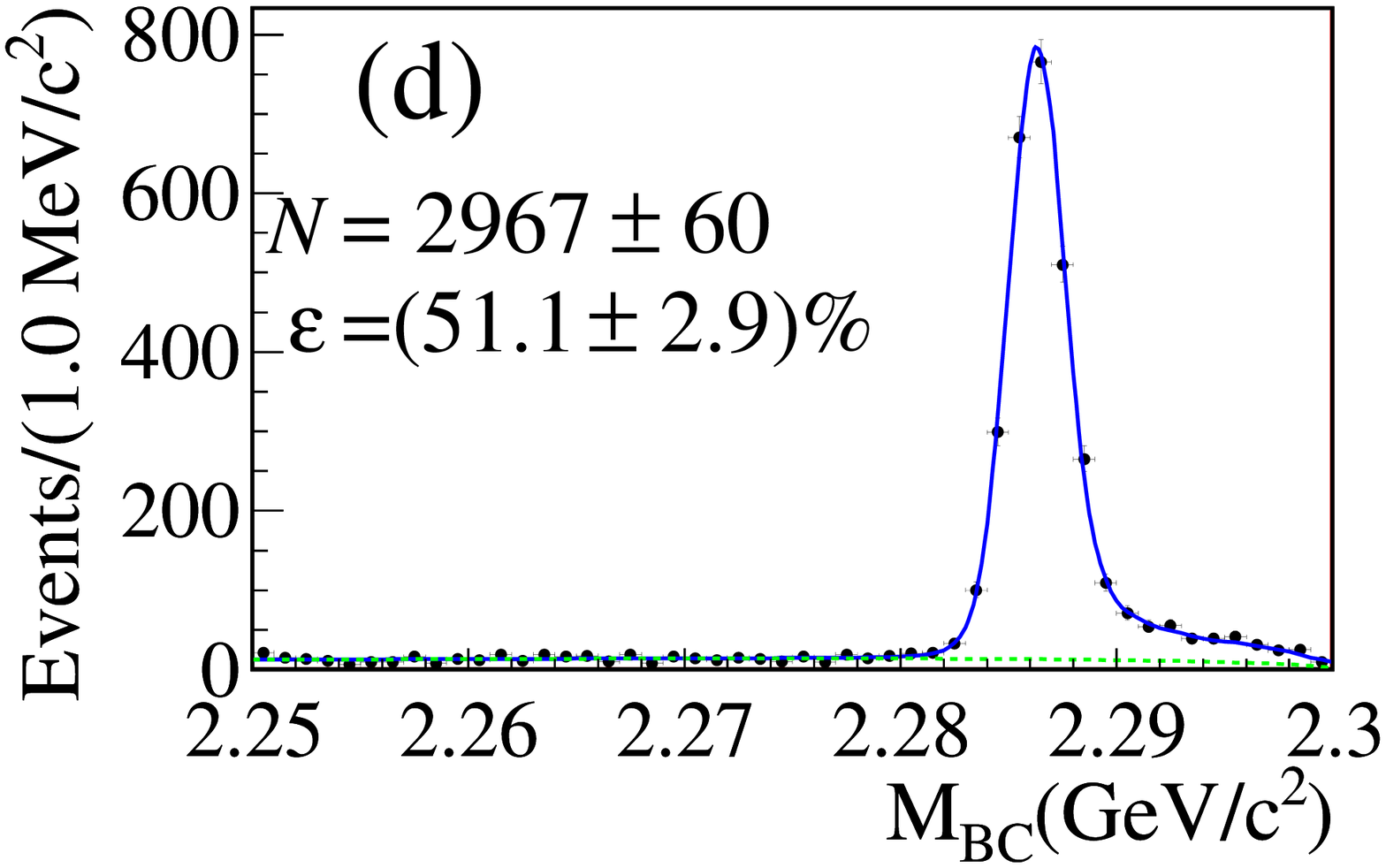}
\end{overpic}
\caption{Fit results of $\mbc$ distribution of $\lamdecay$ in data at $\sqs=\ENERGYAT~\mev$ (a), $\ENERGYBT~\mev$ (b), $\ENERGYCT~\mev$ (c), $\ENERGYDT~\mev$ (d). Dots are Poisson averages of the data in the bins and error bars represent one time of corresponding standard deviations, the blue solid curves are the sum of fit functions, the barely visible green dashed lines are the background shapes. $N$ is the yield with statistical uncertainty of $\lam$ signal, and $\varepsilon$ represents the corresponding detection efficiency and uncertainty.}
\label{4PointsFig}
\end{center}
\end{figure}

According to energy and momentum conservation, two discriminating variables, the \textit{energy difference} $\dele$ and \textit{beam-constrained mass} $\mbc$, are utilized to identify the $\Lam$ signals. The energy difference is defined as $\dele~\equiv~E~-~\ebm$, where $E$ is the energy of the $\Lam$ candidate and $\ebm$ the mean energy of the two colliding beams. In each tagged mode, the $\Lam$ candidates are formed by all possible combinations of the final state particles, and only the one with minimum $|\dele|$ is stored. In the following analysis, events are rejected if they fail the $\dele$ requirements specified in Ref.~\cite{BAM0162}. The beam-constrained mass is defined as $\mbc c^{2}\equiv\sqrt{\ebm^{2}-p^{2}c^{2}}$, where $p$ is the momentum of the $\Lam$ candidate. Both $\dele$ and $\mbc$ are calculated in the initial $e^{+}e^{-}$ CM system. In Fig.~\ref{4PointsFig}, the $\mbc$ distributions for $\lamdecay$ at the four CM energies are shown. Clear peaks at the nominal $\lam$ mass are observed. Studies of the inclusive MC samples show that the cross feeds among the ten tagged modes are less than 1.5\% and the background shape can be described by the ARGUS function~\cite{Argusf}.

Performing an unbinned maximum likelihood fit to each $\mbc$ distribution gives the corresponding event yields, as partly illustrated in Fig.~\ref{4PointsFig}. The signal shape of the fit is obtained from convolving the $\mbc$ shape of MC simulations with a Gaussian function to compensate a possible resolution difference between data and MC simulations. The background is described by an ARGUS function with the high-end truncation fixed. At $\sqs=\ENERGYDT~\mev$, the parameters of the ARGUS and the Gaussian functions used in the convolution are obtained from the fit. At the remaining CM energies, all parameters obtained at the highest energy, except for the mean of the Gaussian, are used to fix parameters in the new fits. Yields are extracted from the signal region 2276~$\mev$~$<\mbc c^{2}$~$<$~$\ebm$ in each fit. The detection efficiency of each decay mode is evaluated by MC simulations of the $\ee$ process. Figure~\ref{4PointsFig} gives the efficiencies of mode $\modeI$ at the four CM energies.

The cross section of the $i$-th mode is determined using
\begin{equation}
\sigma_{i} = \frac{N_{i}}{\varepsilon_{i}\cdot \mathcal{L}_{\textrm{int}}\cdot f_{\textrm{VP}}\cdot BR_{i}\cdot f_{\textrm{ISR}}},
\label{csformula}
\end{equation}
where $N_{i}$ and $\varepsilon_{i}$ represent the yield and corresponding detection efficiency. The integrated luminosity $\mathcal{L}_{\textrm{int}}$ is taken from Ref.~\cite{LumXYZ,LumRscan}. The vacuum polarization (VP) correction factor $f_{\textrm{VP}}$ is calculated to be 1.055 at all four CM energies~\cite{VPfactor}. The $BR_{i}$ represents the product of branching fractions of the $i$-th $\Lam$ decay mode and its subsequent decay(s). The $f_{\textrm{ISR}}$ is the ISR correction factor derived in Ref.~\cite{ISRfactor} and implemented in {\sc kkmc}. Since the calculation of $f_{\textrm{ISR}}$ requires the cross section line-shape as input, an iterative procedure has been performed.
\begin{table}[!htbp]
\footnotesize
\caption{Summary of the reconstruction related, mode-specific, relative systematic uncertainties of the cross section at $\sqs$~=~$\ENERGYDT~\mev$, quoted in \%.}
\label{SysErrs}
\begin{center}
\begin{tabular}{l c c c c c c c c c c }
\hline
\hline
\multicolumn{1}{  c  }{\multirow{2}{*}{} } &
\multicolumn{1}{  c  }{\multirow{2}{*}{} } &
\multicolumn{1}{  c  }{\multirow{2}{*}{} } &
\multicolumn{1}{  c  }{\multirow{2}{*}{} } &
\multicolumn{1}{  c  }{\multirow{2}{*}{} } &
\multicolumn{1}{  c  }{\multirow{2}{*}{} } &
\multicolumn{1}{  c  }{MC} &
\multicolumn{1}{  c  }{~Signal~} &
\multicolumn{1}{  c  }{~Total~} \\
\multicolumn{1}{  c  }{Source}&
\multicolumn{1}{  c  }{Tracking}&
\multicolumn{1}{  c  }{PID}&
\multicolumn{1}{  c  }{$\kst$}&
\multicolumn{1}{  c  }{$\Lambda$}&
\multicolumn{1}{  c  }{$\pi^{0}$}&
\multicolumn{1}{  c  }{statistic}&
\multicolumn{1}{  c  }{~model~}&
\multicolumn{1}{  c  }{~BR~}& \\
\hline
 $\modeI$     & 3.2 & 4.6  &  --  & --  & --  & 0.2  & --   & 6.0 \\
 $\modeII$    & 1.3 & 0.5  &  1.2 & --  & --  & 0.6  & 0.2  & 5.6 \\
 $\modeIII$   & 1.0 & 1.0  &  --  & 2.5 & --  & 0.8  & 0.5  & 6.2 \\
 $\modeIV$    & 3.0 & 7.6  &  --  & --  & 1.0 & 0.6  & 2.0  & 8.3 \\
 $\modeV$     & 1.0 & 1.8  &  1.2 & --  & 1.0 & 1.1  & 1.0  & 7.5 \\
 $\modeVI$    & 1.0 & 1.0  &  --  & 2.5 & 1.0 & 0.6  & 0.6  & 6.0 \\
 $\modeVII$   & 2.8 & 5.3  &  1.2 & --  & --  & 1.0  & 0.5  & 9.3 \\
 $\modeVIII$  & 3.0 & 3.0  &  --  & 2.5 & --  & 0.9  & 0.8  & 7.9 \\
 $\modeIX$    & 1.0 & 1.0  &  --  & 2.5 & --  & 1.1  & 1.7  & 6.7 \\
 $\modeX$     & 3.0 & 4.0  &  --  & --  & 1.0 & 0.8  & 0.8  & 7.4 \\
\hline
\hline
\end{tabular}
\end{center}
\end{table}

The systematic uncertainties of the cross section can be classified into \textit{reconstruction related} and \textit{general} contributions. The reconstruction related contributions are mode-specific and mainly originate from tracking, PID, reconstruction of intermediate states and total BRs. The uncertainties of $\dele$ and $\mbc$ requirements are negligible after correcting for the difference in resolution between simulated and real data samples. The uncertainties from tracking and PID of charged particles are investigated using control samples from $e^{+}e^{-}~\rightarrow~\pi^{+}\pi^{-}\pi^{+}\pi^{-}$, $K^{+}K^{-}\pi^{+}\pi^{-}$ and $p\bar{p}\pi^{+}\pi^{-}$ collected at $\sqs > 4.0~\gev$~\cite{BAM0216}. The uncertainties are obtained after weighting according to the momenta of the corresponding final states. Reconstruction uncertainties of $\kst$, $\Lambda$ and $\pi^{0}$ have been found to be 1.2\%, 2.5\% and 1.0\%~\cite{BAM0162}. Statistical uncertainties of detection efficiencies are considered as systematic uncertainties. The dependence of the reconstruction efficiency on the MC model for the ten decay modes also gives a small contribution to the systematic uncertainty~\cite{BAM0162}. Uncertainties originating from the total BRs of the tagged modes are quoted from Refs.~\cite{BAM0162,PDG}. A summary of the reconstruction related systematic uncertainties are given in Table~\ref{SysErrs}. The total uncertainty at each energy has been calculated assuming that the values given at $\sqs$~=~$\ENERGYDT$ $\mev$ are valid at all CM energies.
\begin{table}[!htbp]
\caption{Summary of the general relative systematic uncertainties of the cross section originating from the factors $f_{\textrm{ISR}}$, $f_{\textrm{VP}}$ and $\mathcal{L}_{\textrm{int}}$, quoted in \%.}
\footnotesize
\label{SysErrsII}
\begin{center}
\begin{tabular}{ l  c  c  c  c  c  c  c }
\hline
\hline
\multicolumn{1}{  c  }{\multirow{3}{*}{}} &
\multicolumn{5}{  c  }{$f_{\textrm{ISR}}$} &
\multicolumn{1}{  c  }{\multirow{3}{*}{}} &
\multicolumn{1}{  c  }{\multirow{3}{*}{}} \\ \cline{2-6}
\multicolumn{1}{  c  }{}&
\multicolumn{1}{  c  }{Calculation}&
\multicolumn{1}{  c  }{Line-}&
\multicolumn{1}{  c  }{CM} &
\multicolumn{1}{  c  }{Energy} &
\multicolumn{1}{  c  }{}&
\multicolumn{1}{  c  }{}&
\multicolumn{1}{  c  }{} \\
\multicolumn{1}{  c  }{$\sqs$ ($\mev$)}&
\multicolumn{1}{  c  }{model} & shape & energy & Spread  & Total &
\multicolumn{1}{  c  }{~~$f_{\textrm{VP}}$}&
\multicolumn{1}{  c  }{~$\mathcal{L}_{\textrm{int}}$~} \\ \cline{1-8}
\multicolumn{1}{  c  }{\ENERGYAT} &
\multicolumn{1}{  c   }{3.4} & 1.2 & 18.0 & 3.0 & 18.6 & ~~0.5  &  1.0  \\
\multicolumn{1}{  c  }{\ENERGYBT} &
\multicolumn{1}{  c  }{0.7 } &  0.6 & -- & 0.2 & 0.9 & ~~0.5  &  0.7  \\
\multicolumn{1}{  c  }{\ENERGYCT} &
\multicolumn{1}{  c  }{0.2 } & 1.7 & -- & -- & 1.7 & ~~0.5  &  0.7  \\
\multicolumn{1}{  c  }{\ENERGYDT} &
\multicolumn{1}{  c  }{0.1 } & 2.6 & -- & -- & 2.6 & ~~0.5  &  1.0  \\
\hline
\hline
\end{tabular}
\end{center}
\end{table}

The general contributions to the systematic uncertainty originate from uncertainties in $f_{\textrm{ISR}}$, $f_{\textrm{VP}}$ and $\mathcal{L}_{\textrm{int}}$ in Eq.~(\ref{csformula}) and are the same for all decay modes. 
The $f_{\textrm{ISR}}$ is obtained using the {\sc kkmc} generator which requires a cross section line-shape as input. The line-shape is in turn obtained by an iterative fitting procedure of the cross section data using Eq.~\eqref{OnePhotonPre}. In the fit, the $\ratios$ value at an arbitrary CM energy is assigned by linear interpolation between the two known values listed in Table~\ref{AngDisParas}. For simplicity, $|G_{M}|$ is assumed to be independent of the CM energy. To precisely describe the data, the $\alpha$ in the Sommerfeld resummation factor is replaced by $\alpha_{s}(=0.25)$. In the line-shape, the cross section at the CM energy region  $(2m_{\Lam}c^{2},~\ENERGYAT)$~$\mev$ is obtained from extrapolating the fit; below threshold it vanishes, as shown by the blue solid curve in Fig.~\ref{FitLinShp}. Four sources of systematic uncertainty from the $f_{\textrm{ISR}}$ are considered: First, the uncertainty of the \textit{calculation model} is studied using a different algorithm mentioned in Ref.~\cite{ConExc}. Second, the uncertainty associated with the input \textit{line-shape} is estimated using different fit functions. Third, the $f_{\textrm{ISR}}$ depends on the CM energy of the $\ee$ process. The uncertainty of the \textit{CM energy} therefore contributes near the threshold. At the lowest energy point, the CM energy is measured to be $\sqs=4574.50\pm0.72$~$\mev$~\cite{XYZECMS}. Finally, the beam \textit{energy spread}, which has been estimated as  $1.55\pm0.18~\mev$, is important near threshold and contributes to the $f_{\textrm{ISR}}$ uncertainty. For the other, higher, energies, the effects from the CM energy uncertainty and the beam energy spread are less than 0.1\% and can be neglected due to the flat line-shape of the cross section. The uncertainty of $f_{\textrm{VP}}$ is calculated to be 0.5\% at all four CM energies~\cite{VPfactor}. The uncertainty from the integrated luminosity has been found to be 0.7\% at $\sqs$~=~$\ENERGYBT$ and $\ENERGYCT$~$\mev$ and 1.0\% at $\sqs$~=~$\ENERGYAT$ and $\ENERGYDT~\mev$~\cite{LumXYZ,LumRscan}.  A summary of the general contributions to the systematic uncertainties is given in Table~\ref{SysErrsII}.

The cross sections obtained in different decay modes are combined using the method mentioned in Ref.~\cite{AverageData}, in which the cross section is given by:
\begin{equation}
\sigma =\sum_{i}w_{i}\sigma_{i}~~~\textmd{with}~~~w_{i}=(1/\Delta\sigma_{i}^{2})\bigg/\bigg(\sum_{i}1/\Delta\sigma_{i}^{2}\bigg).
\label{averagex}
\end{equation}
Here, $w_{i}$ and $\Delta\sigma_{i}$ denote the weight and the total uncertainty, respectively, of the measured cross section $\sigma_{i}$ of mode $i$. The sum is performed over all 20 decay modes of $\lam$ and $\lambar$~\cite{IndividualCSs}. 
The combined uncertainty is calculated by
\begin{equation}
\Delta\sigma^{2} = \sum_{i,j}w_{i}(\mathbf{M}_{\sigma})_{ij}w_{j},
\label{variance}
\end{equation}
where $\mathbf{M}_{\sigma}$ represents the covariance matrix of these cross section measurements, in which the correlations between any two measurements $\sigma_{i}$ and $\sigma_{j}$ are considered. The resulting cross sections at the four CM energies are listed in Table~\ref{CroSecSum} and shown in Fig.~\ref{FitLinShp} together with the Belle data~\cite{Belle} for comparison.
\begin{table}[!htbp]
\caption{The average cross section of $\ee$ measured at each CM energy, where the uncertainties are statistical and systematic, respectively. The observed cross section can be obtained by multiplying the $f_{\textrm{ISR}}$ and the $\sigma$.}
\label{CroSecSum}
\begin{center}
\begin{tabular}{ c  c  c  c}
\hline
\hline
\multicolumn{1}{  c  }{\multirow{2}{*}{~~$\sqs$ ($\mev$)~~} }&
\multicolumn{1}{  c  }{\multirow{2}{*}{~~$\mathcal{L}_{\textrm{int}}$ $(\textmd{pb}^{-1})$~~} }&
\multicolumn{1}{  c  }{\multirow{2}{*}{~~$f_{\textrm{ISR}}$~~} } &
\multicolumn{1}{  c  }{\multirow{2}{*}{~~~~~~~~~$\sigma$ (pb)~~~~~~~~~} } \\ 
\multicolumn{1}{  c   }{}&
\multicolumn{1}{  c   }{}&
\multicolumn{1}{  c   }{}&
\multicolumn{1}{  c   }{} \\ \cline{1-4}
 $\ENERGYAT$   & 47.67   & 0.45  &  $\BornCSA$  \\
 $\ENERGYBT$   & 8.54   & 0.66  &  $\BornCSB$  \\
 $\ENERGYCT$   & 8.16   & 0.71  &  $\BornCSC$  \\
 $\ENERGYDT$   & 566.93   & 0.74  &  $\BornCSD$  \\
\hline
\hline
\end{tabular}
\end{center}
\end{table}

\begin{figure}[!htbp]
\setlength{\abovecaptionskip}{0.0cm}
\setlength{\belowcaptionskip}{-0.5cm}
\begin{center}
\includegraphics[width=3.33in,height=2.22in,angle=0]{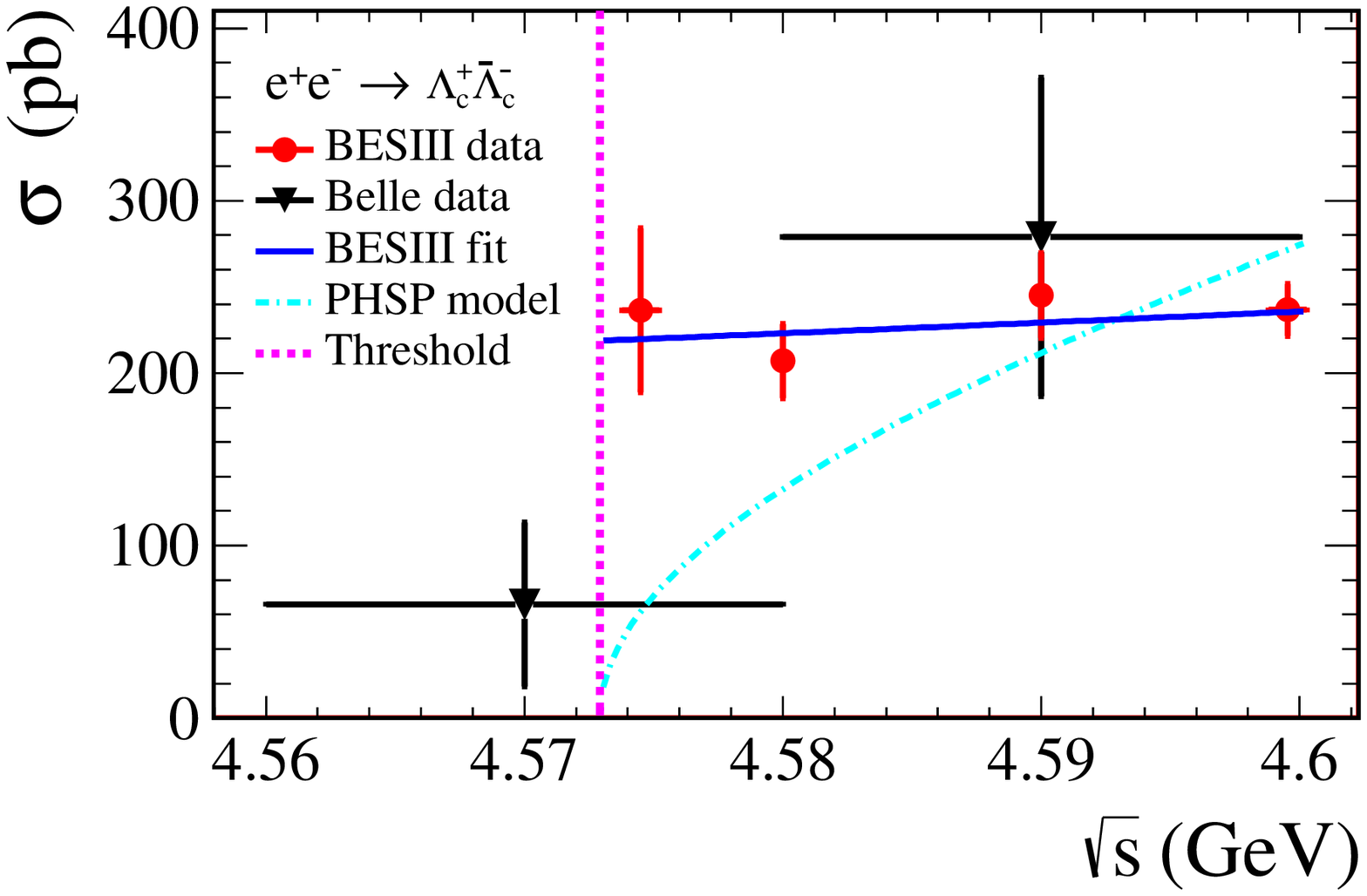}
\caption{ Cross section of $\ee$ obtained by BESIII (this work) and Belle. The blue solid curve represents the input line-shape for {\sc kkmc} when determining the $f_{\textrm{ISR}}$. The dash-dot cyan curve denotes the prediction of the phase space (PHSP) model, which is parameterized by Eq.~\ref{OnePhotonPre} but with $C=1$ and flat $|G_{M}|$ with respect to $\sqs$.}
\label{FitLinShp}
\end{center}
\end{figure}

\begin{figure}[!htbp]
\setlength{\abovecaptionskip}{0.2cm}
\setlength{\belowcaptionskip}{-0.5cm}
\begin{center}
\begin{overpic}[width=1.68in,height=1.12in,angle=0]{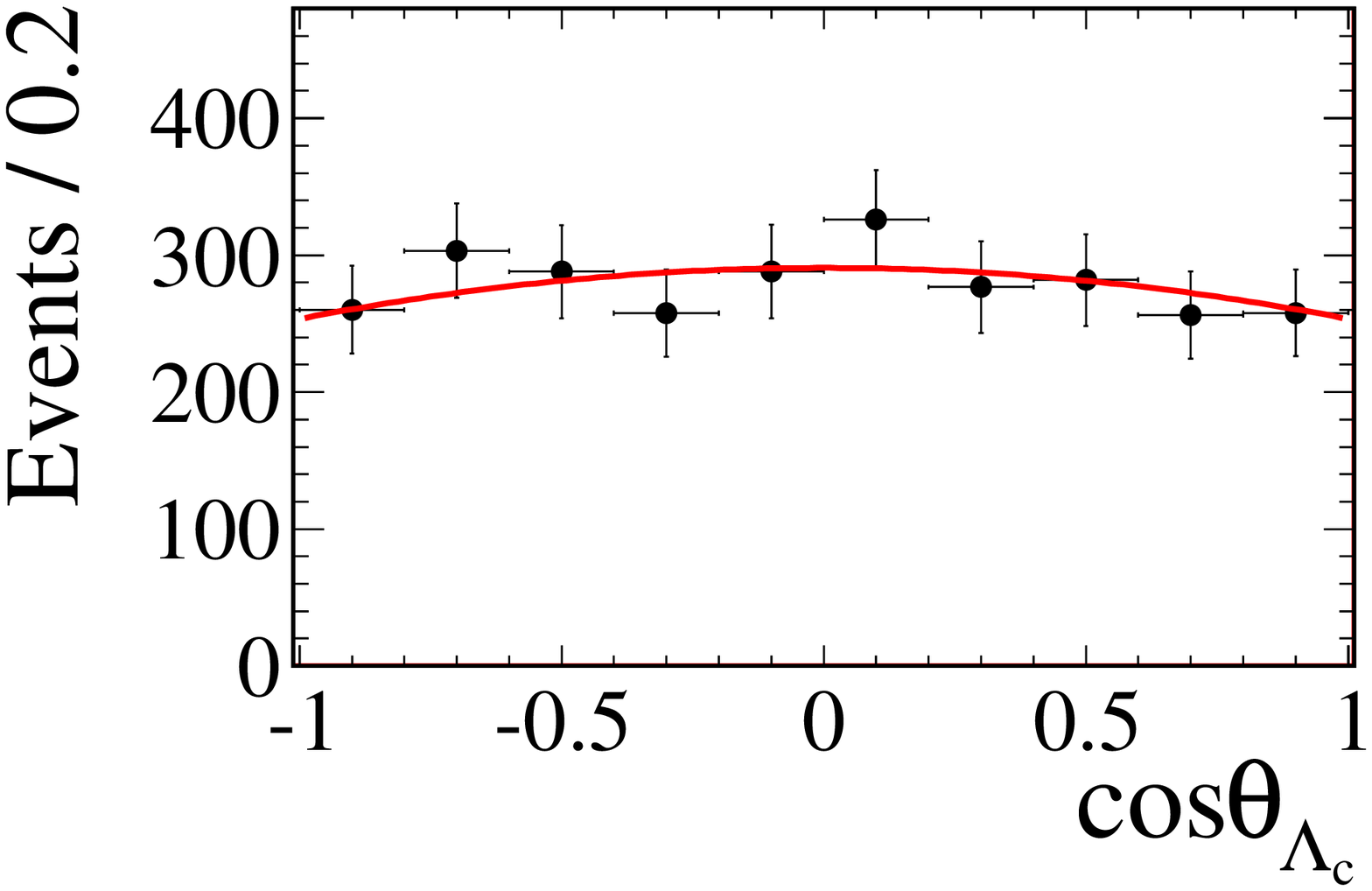}
\end{overpic}
\begin{overpic}[width=1.68in,height=1.12in,angle=0]{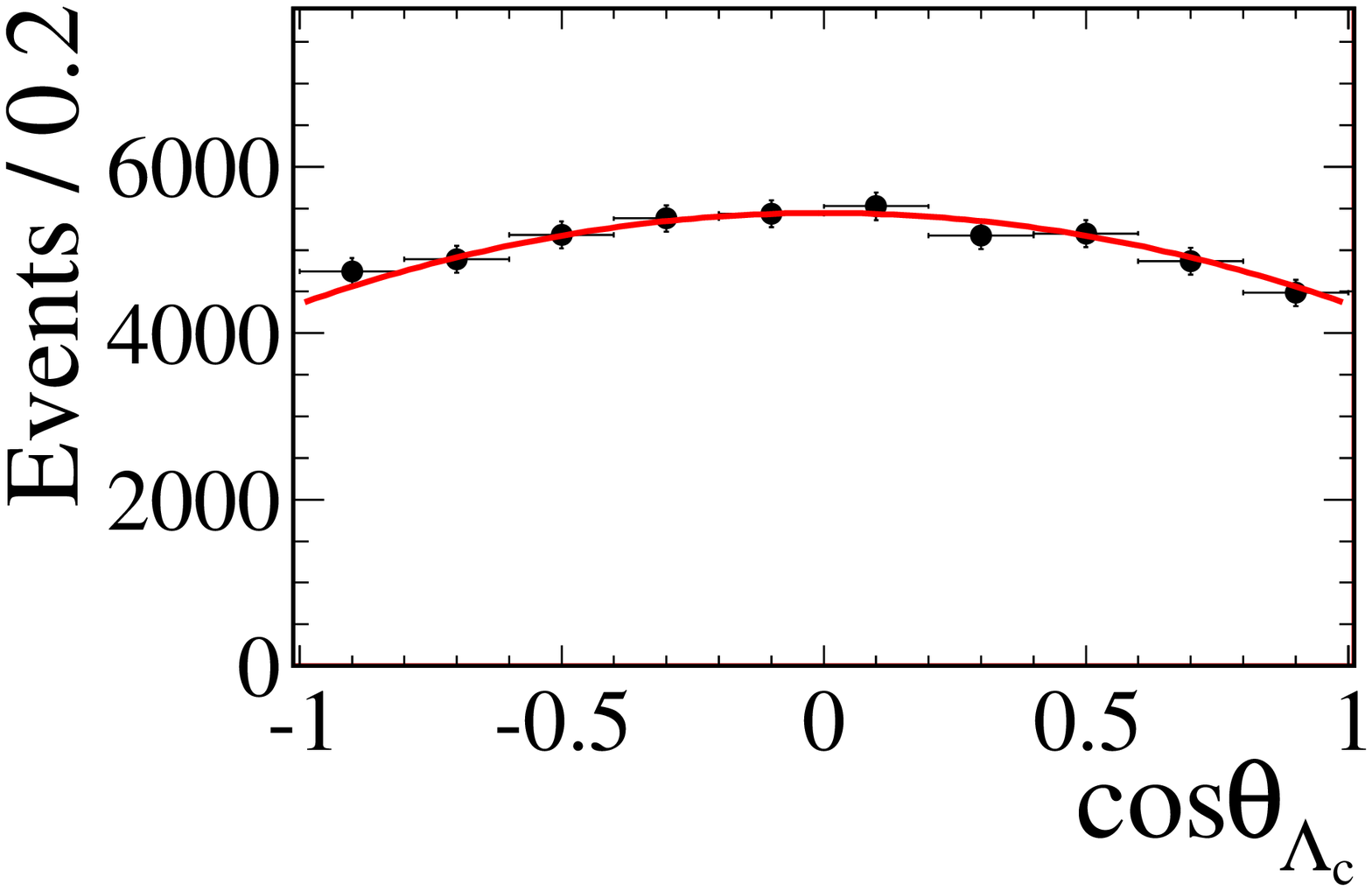}
\end{overpic}
\caption{Angular distribution after efficiency correction and results of the fit to data at $\sqs$~=~$\ENERGYAT$~$\mev$ (left) and $\ENERGYDT$ $\mev$ (right). }
\label{AngDisFit}
\end{center}
\end{figure}

The data sets collected at $\sqs$~=~$\ENERGYAT$ and $\ENERGYDT$~$\mev$ are large enough to perform a detailed study in the CM frame of the $\Lam$ polar angle $\theta_{\Lam}$, which is defined as the angle between the $\Lam$ momentum and the beam direction. The data fulfilling all selection criteria are divided into ten bins in $\cos\theta_{\lam}$. In each $\cos\theta_{\lam}$ bin, the total yield is obtained by summing the yields of all the ten tagged modes. The one-dimensional bin-by-bin efficiency corrections are applied on these total yields. The same procedure is performed by tagging $\lambar$ decay channels. The total yields of $\lam$ and $\lambar$ are combined bin-by-bin, and the shape function $f(\theta)~\propto~(1+\alpha_{\Lam} \cos^{2}\theta)$ is fitted to the combined data, as shown in Fig.~\ref{AngDisFit}. Table~\ref{AngDisParas} lists the resulting $\alpha_{\Lam}$ parameters obtained from the fits, as well as the $\ratios$ ratios extracted using the equation
\begin{equation}
\label{ratios}
\ratios^{2}(1-\beta^{2})=(1-\alpha_{\Lam})/(1+\alpha_{\Lam}).
\end{equation}

\begin{table}[!htbp]
\caption{Shape parameters of the angular distribution and $\ratios$ ratios at $\sqs$~=~$\ENERGYAT$ and $\ENERGYDT$~$\mev$. The uncertainties are statistical and systematic, respectively. }
\label{AngDisParas}
\begin{center}
\begin{tabular}{c   c  c }
\hline
\hline
~$\sqs$ ($\mev$)~~ & ~~~~~~~~~~~~$\alpha_{\Lam}$~~~~~~~~~~ & ~~~~~~~~~~$\ratios$~~~~~~~~~~\\
\hline
\ENERGYAT  & $\alpHaA$   & $\RatioA$\\
\ENERGYDT  & $\alpHaD$   & $\RatioD$\\
\hline
\hline
\end{tabular}
\end{center}
\end{table}

The systematic uncertainties of the $\alpha_{\Lam}$ considered here are the contributions from the fit range and the bin size. A change of the fit range in $\cos\theta$ from ($-1.0,~1.0$) to ($-0.8,~0.8$) and in the number of bins from 10 to 20 are performed, and the differences in the obtained $\alpha_{\Lam}$ are regarded as the systematic uncertainty. Systematics originating from the model dependencies in the efficiency correction are found to be negligible compared to the statistical uncertainties.

In summary, using data collected at $\sqs=\ENERGYAT$, $\ENERGYBT$, $\ENERGYCT$, and $\ENERGYDT~\mev$ with the BESIII detector, the cross sections of $\ee$ have been measured with high precision, by reconstructing $\lam$ and $\lambar$ independently with ten Cabibbo-favored hadronic decay channels. The most precise cross section measurement is achieved so far at $\sqs=\ENERGYAT~\mev$, which is only $1.6~\mev$ above the threshold. The measured value is ($\BornCSA$)~pb, which highlights the enhanced cross section near threshold and indicates the complexity of production behavior of the $\Lam$. At $\sqs=\ENERGYAT$ and $\ENERGYDT~\mev$, the data samples are large enough to study polar angle distributions of $\Lam$ and measure the $\Lam$ form factor ratio $\ratios$ for the first time. These results provide important insights into the production mechanism and structure of the $\Lam$ baryons.

The BESIII collaboration thanks the staff of BEPCII, the IHEP computing center and the supercomputing center of USTC for their strong support. This work is supported in part by National Key Basic Research Program of China under Contract No. 2015CB856700; National Natural Science Foundation of China (NSFC) under Contracts Nos. 11235011, 11335008, 11375205, 11425524, 11625523, 11635010; the Chinese Academy of Sciences (CAS) Large-Scale Scientific Facility Program; the CAS Center for Excellence in Particle Physics (CCEPP); Joint Large-Scale Scientific Facility Funds of the NSFC and CAS under Contracts Nos. U1332201, U1532257, U1532258; CAS under Contracts Nos. KJCX2-YW-N29, KJCX2-YW-N45, QYZDJ-SSW-SLH003; 100 Talents Program of CAS; National 1000 Talents Program of China; INPAC and Shanghai Key Laboratory for Particle Physics and Cosmology; German Research Foundation DFG under Contracts Nos. Collaborative Research Center CRC 1044, FOR 2359; Istituto Nazionale di Fisica Nucleare, Italy; Koninklijke Nederlandse Akademie van Wetenschappen (KNAW) under Contract No. 530-4CDP03; Ministry of Development of Turkey under Contract No. DPT2006K-120470; National Natural Science Foundation of China (NSFC) under Contracts Nos. 11505034, 11575077; National Science and Technology fund; The Swedish Research Council; U. S. Department of Energy under Contracts Nos. DE-FG02-05ER41374, DE-SC-0010118, DE-SC-0010504, DE-SC-0012069; University of Groningen (RuG) and the Helmholtzzentrum fuer Schwerionenforschung GmbH (GSI), Darmstadt; WCU Program of National Research Foundation of Korea under Contract No. R32-2008-000-10155-0




\begin{thebibliography}{99}
\bibitem{IndividualCSs}
At $\sqs$~=~$\ENERGYDT$~$\mev$, the deviations between any two of the 20 individual Born cross section can be covered by corresponding total uncertainty. While it is not the case for a few decay channels at other three energies due to the low statistics of the data sample.
\end{thebibliography}
\end{document}